\documentclass[aps,prl,twocolumn,superscriptaddress,showpacs]{revtex4-1}
\usepackage{amsmath,amssymb,graphics,epsfig,epstopdf,color,verbatim,tabularx}
\usepackage[colorlinks,linkcolor=blue,citecolor=blue,urlcolor=blue]{hyperref}

\begin{document}

\title{The amplitude mode in three-dimensional dimerized antiferromagnets}

\author{Yan Qi Qin}
\affiliation{Beijing National Laboratory for Condensed Matter Physics and 
\\ Institute of Physics, Chinese Academy of Sciences, Beijing 100190, 
China}

\author{B. Normand}
\affiliation{Laboratory for Neutron Scattering and Imaging, Paul Scherrer 
Institute, CH-5232 Villigen PSI, Switzerland}

\author{Anders W. Sandvik}
\affiliation{Department of Physics, Boston University, 590 Commonwealth 
Avenue, Boston, Massachusetts 02215, USA}

\author{Zi Yang Meng}
\affiliation{Beijing National Laboratory for Condensed Matter Physics and 
\\ Institute of Physics, Chinese Academy of Sciences, Beijing 100190, China}

\begin{abstract}
The amplitude (``Higgs'') mode is a ubiquitous collective excitation related 
to spontaneous breaking of a continuous symmetry. We combine quantum Monte 
Carlo (QMC) simulations with stochastic analytic continuation to investigate 
the dynamics of the amplitude mode in a three-dimensional dimerized quantum 
spin system. We characterize this mode by calculating the spin and dimer 
spectral functions on both sides of the quantum critical point, finding 
that both the energies and the intrinsic widths of the excitations satisfy 
field-theoretical scaling predictions. While the line width of the spin 
response is close to that observed in neutron scattering experiments on 
TlCuCl$_3$, the dimer response is significantly broader. Our results 
demonstrate that highly non-trivial dynamical properties are accessible 
by modern QMC and analytic continuation methods.
\end{abstract}

\date{\today} 

\maketitle

The spontaneous breaking of a continuous symmetry allows collective 
excitations of the direction and amplitude of the order parameter; 
for O($N$) symmetry, there are $N \! - \! 1$ massless directional 
(Goldstone) modes and one massive amplitude mode \cite{Goldstone1962,
Higgs1964,Zinn-Justin2002,Sachdev2011}. In loose analogy with the Standard 
Model, the latter is often called a Higgs mode. A strongly damped 
amplitude mode has been reported in two dimensions (2D) at the Mott 
transition of ultracold bosons \cite{Endres2012} and at the disorder-driven 
superconductor--insulator transition \cite{Swanson2014,Sherman2015}. In 3D, 
the amplitude mode is expected on theoretical grounds to be more robust, 
and indeed the cleanest observation to date of a ``Higgs boson'' in 
condensed matter is at the pressure-induced magnetic quantum phase 
transition (QPT) in the dimerized quantum antiferromagnet TlCuCl$_3$ 
\cite{Ruegg2004,Ruegg2008,Merchant2014}. 

Below the upper critical number of space-time dimensions, which for an 
O($N$) model is $D_c = 4$, the amplitude mode is unstable, decaying 
primarily into pairs of Goldstone bosons \cite{Sachdev1999,Zwerger2004,
Dupuis2011}. In both 2D and 3D, the longitudinal dynamic susceptibility 
exhibits an infrared singularity due to the Goldstone modes 
\cite{Podolsky2011}, whose consequences for the visibility of the amplitude 
mode have been investigated extensively in 2D \cite{Podolsky2012,Gazit2013,
GazitPRB2013}. It was noted \cite{Podolsky2011} that the scalar 
O($N$)-symmetric susceptibility remains uncontaminated by infrared 
contributions, which should permit the amplitude mode to be observed 
as a well-defined peak. The (3+1)D O(3) case of TlCuCl$_3$ is at $D_c$ 
and the amplitude mode is critically damped, meaning that its width is 
proportional to its energy at the mean-field level \cite{Ruegg2008,ra,raw,
Kulik11}. This mode can be probed through the spin response (longitudinal 
susceptibility) by neutron spectroscopy, and measurements over a wide range 
of pressures reveal a rather narrow peak width of just 15\% of the excitation 
energy \cite{Merchant2014}. The value of this near-constant width-to-energy 
ratio is the key to the mode visibility, thus calling for unbiased numerical 
calculations in suitable model Hamiltonians.

\begin{figure}[t]
\includegraphics[width=7.0cm]{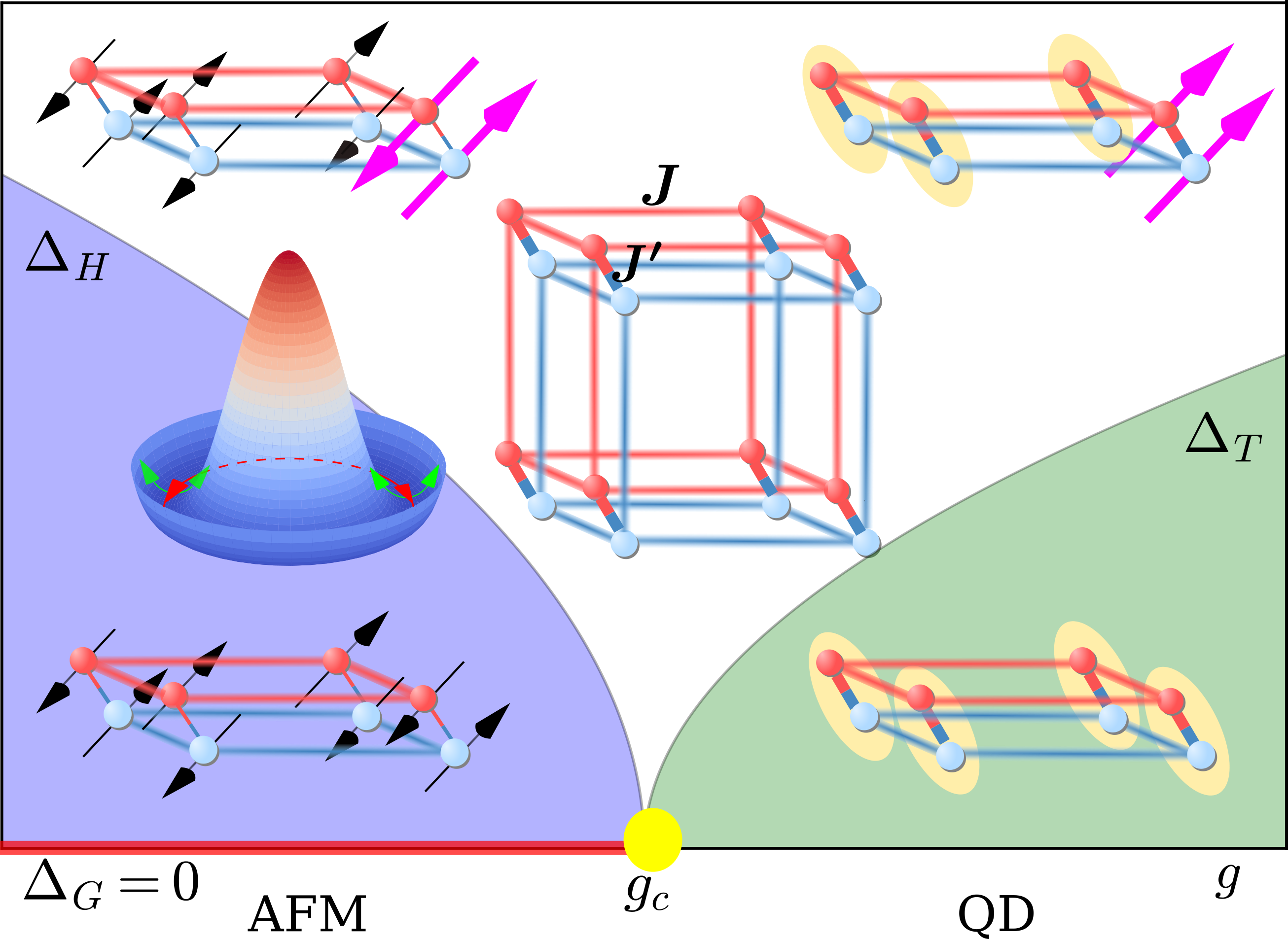}
\caption{Schematic representation of ground states, excitation processes, and 
corresponding gaps in a dimerized antiferromagnet. The ratio $g = J'/J$ of the 
intra- and inter-dimer coupling constants controls a QPT from an AFM to a QD 
state. In the AFM phase, the excitations are two gapless spin waves (Goldstone 
modes, $\Delta_G = 0$, red line) plus an amplitude mode with gap $\Delta_H$, 
corresponding respectively to axial and radial fluctuations in the ``Mexican 
hat'' potential. In the QD phase, singlet--triplet dimer excitations have gap 
$\Delta_T$.}
\label{fig:gap}
\end{figure}

In this Letter, we provide a systematic investigation of the dynamics and 
scaling of the amplitude mode at coupling values across the QPT in a 3D 
dimerized spin-$1/2$ antiferromagnet, by performing large-scale stochastic 
series expansion quantum Monte Carlo (SSE-QMC) simulations and applying 
advanced stochastic analytic-continuation (SAC) methods. Thus we provide 
an unbiased numerical demonstration that the amplitude mode is 
critically damped and that its energy, width, and height obey 
field-theoretical predictions. Beyond these universal scaling forms, we 
quantify the nonuniversal width-to-energy ratios of the amplitude-mode 
peaks in the spin and dimer channels.

We consider the double-cubic geometry shown in Fig.~\ref{fig:gap} 
\cite{Jin2012}, which consists of two simple cubic lattices whose sites 
are connected pairwise by nearest-neighbor Heisenberg exchange interactions, 
$J_{ij} {\vec S}_i \! \cdot \! {\vec S}_j$, with $J_{ij} = J$ in each cubic 
lattice and $J_{ij} = J'$ for inter-cube (dimer) bonds. Increasing the 
ratio $g \! = \! J'/J$ drives a QPT where the ground state changes from a 
``renormalized classical'' \cite{rchn,rcsy} antiferromagnetic (AFM) state 
to a quantum disordered (QD) dimer-singlet state (Fig.~\ref{fig:gap}). This 
transition is in the same universality class as the pressure-driven QPT in 
TlCuCl$_3$. In a recent QMC analysis of the static properties of the 
double-cubic system \cite{Qin2015}, we established the quantum critical 
point (QCP) as $g_c = 4.83704(6)$ and quantified the logarithmic (log) 
scaling corrections expected near criticality in the AFM state at $D_c$. 

We use SSE-QMC \cite{Sandvik1999,Syljuaasen2002} to measure both spin and 
dimer correlation functions in imaginary time; technical details may be 
found in Sec.~SI of the Supplemental Material (SM \cite{sm}). The 
former probes $S = 1$ excitations of the ground state and contains the 
longitudinal susceptibility, while the latter, the symmetric scalar response 
\cite{Podolsky2011,Podolsky2012,Gazit2013}, probes $S = 0$ excitations. 
We employ SAC methods \cite{Sandvik1998,Beach2004,Syljuasen2008,Fuchs2010,
Sandvik2015,Shao2016} to obtain high-resolution data for the spin and dimer 
spectral functions, and discuss the concepts and practicalities of this 
procedure in Sec.~SII of the SM \cite{sm}. Depending on the value of $g$, 
both spectral functions contain features arising from the Goldstone, amplitude, 
and triplon (gapped singlet-triplet) excitations. Henceforth we use the term 
``Higgs'' as shorthand for the amplitude-mode contributions. The nature and 
energies of these modes are represented schematically in Fig.~\ref{fig:gap}. 

Our simulations are performed on a system of $N = 2L^3$ sites at an inverse 
temperature $J\beta = 2L$, such that the low-temperature limit, $T \to 0$, 
is achieved as $L \to \infty$. The dynamical magnetic ($S = 1$) response is 
obtained from the spin correlation function
\begin{equation}
\label{eq:spin_cor}
S(\mathbf{q},\tau) = \langle S^z_{-\mathbf{q}}(\tau) S^z_{\mathbf{q}}(0) \rangle,
\end{equation}
where $\tau$ is the imaginary time [Eq.~(S1)] and
\begin{equation}
S^z_{\mathbf{q}} = \frac{1}{\sqrt{N}} \mbox{$\sum_{\mathbf{r}}$} {\rm e}^{-i\mathbf{q}
\cdot\mathbf{r}}(S^{1z}_{r} - S^{2z}_{r}),
\end{equation}
where superscripts 1 and 2 
denote the two cubic lattices. When analytically continued to real frequency, 
$S(\mathbf{q},\tau)$ gives the dynamical structure factor, $S(\mathbf{q},
\omega)$, measured by inelastic neutron scattering. Our simulations contain no 
breaking of spin-rotation symmetry and thus do not separate the longitudinal 
and transverse components of $S(\mathbf{q},\omega)$ explicitly. The Higgs mode 
of the AFM phase is contained in the longitudinal part, but the transverse 
part contains both spin-wave excitations and a multimagnon continuum that 
could obscure the Higgs contribution in the rotationally averaged 
$S(\mathbf{q},\omega)$. However, unlike the 2D case \cite{Sandvik2001}, the 
transverse continuum is expected to be very small in 3D, especially at the 
staggered wave vector, $\mathbf{q} = \mathbf{Q} = (\pi,\pi,\pi)$, on which 
we focus here. 

The scalar ($S = 0$) dynamical response is obtained from the dimer correlation 
function at the zone center, $\mathbf{q} = \mathbf{\Gamma} = (0,0,0)$,
which is given by 
\begin{equation}
\label{eq:dimer_cor}
D(\mathbf{\Gamma},\tau) = \langle B_{\mathbf{\Gamma}}(\tau) B_{\mathbf{\Gamma}}(0) 
\rangle,~~B_{\mathbf{\Gamma}} = \frac{1}{\sqrt{N}} \sum_r B_r,
\end{equation}
where $B_r = \mathbf{S}_{r}^{1} \cdot \mathbf{S}_{r}^{2} - \langle \mathbf{S}_{r}
^{1} \cdot \mathbf{S}_{r}^{2}\rangle$ is the inter-cubic dimer bond operator. 
This quantity was also employed in a recent study of the (2+1)D (bilayer) 
model \cite{Wessel2015}. The real-frequency quantity $D(\mathbf{\Gamma},
\omega)$ may be probed experimentally by Raman scattering \cite{rfl,rss}.

\begin{figure}[t]
\vskip-1mm
\includegraphics[width=7.00cm]{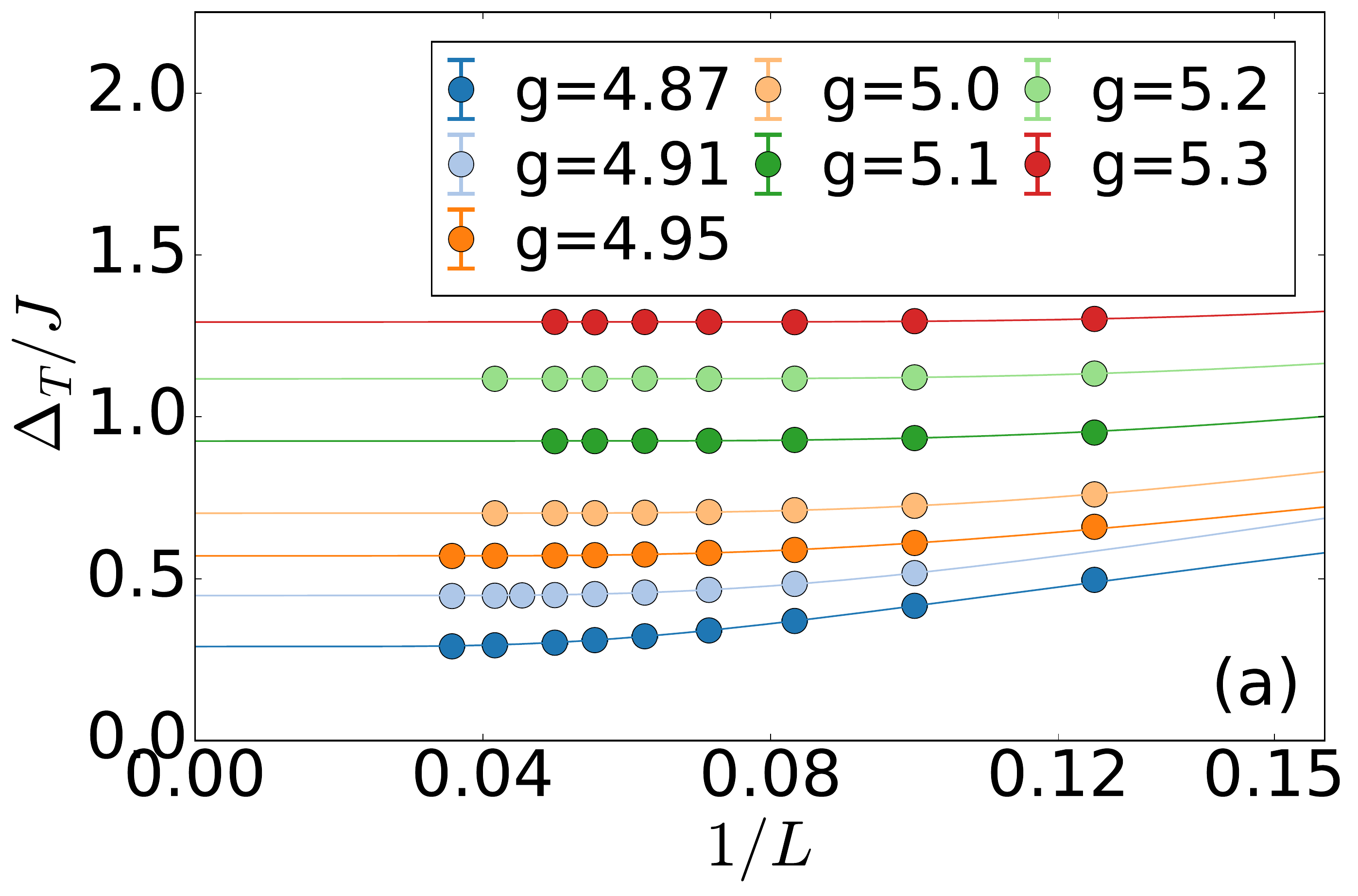}
\includegraphics[width=7.15cm]{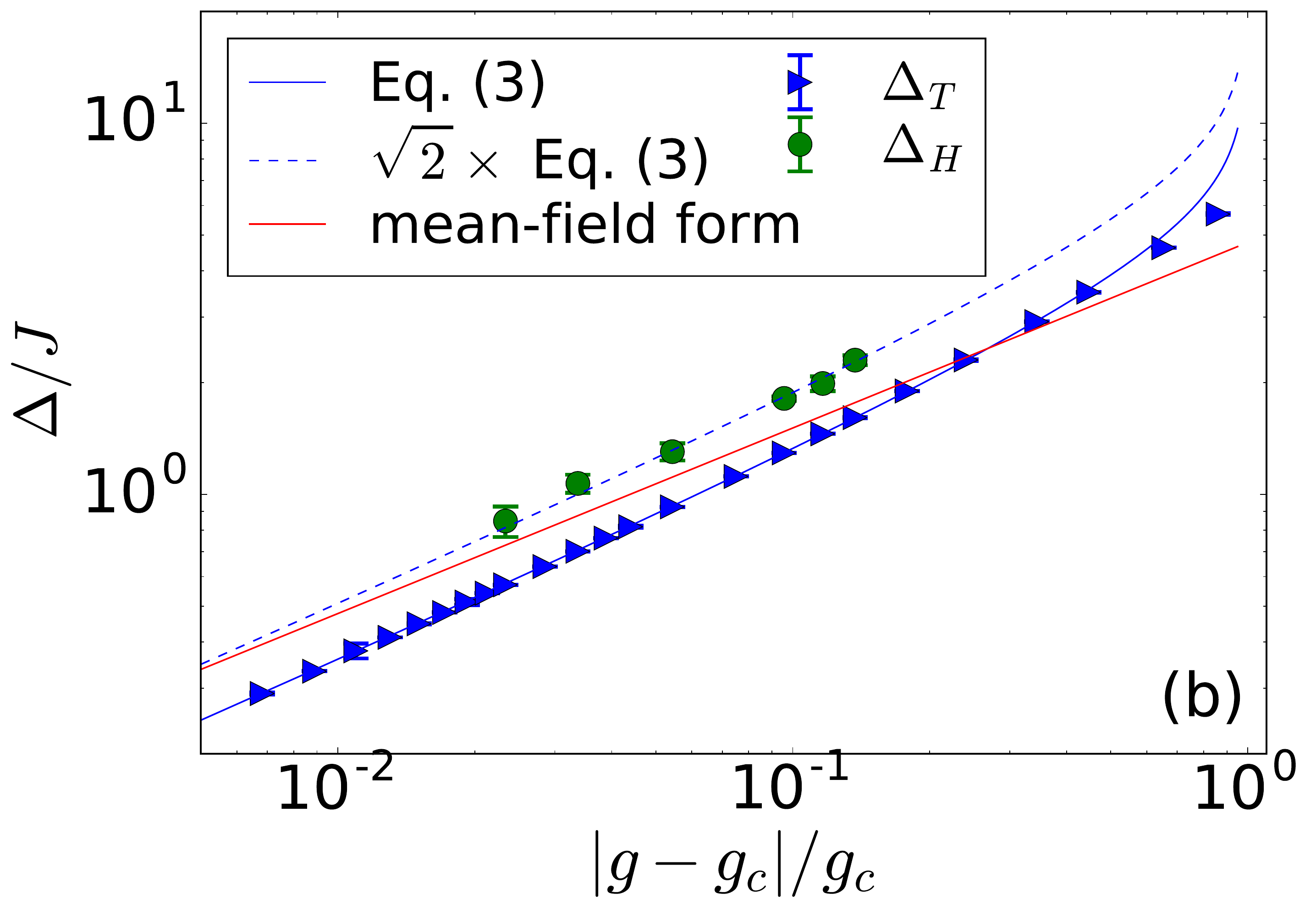}
\caption{(a) Extrapolation of finite-size triplon gaps, using the form 
$\Delta_T(L) = a \exp(-bL) + c$, shown for selected values of $g > g_c$ 
(QD phase). (b) Triplon gaps in the thermodynamic limit (blue triangles), 
shown as a function of $|g - g_c|/g_c$. The red line is a pure mean-field 
(square-root) form, the blue line includes the log correction of 
Eq.~(\ref{eq:lst}) with fitted exponent $\hat{\nu} = 0.230(2)$, and green 
points show the extrapolated Higgs energy, $\Delta_H$, obtained for values 
of $g < g_c$ (AFM phase) mirroring those used in panel (a). The blue dashed 
line is the log-corrected $\Delta_T$ result multiplied by $\sqrt{2}$. Error 
bars in both panels are smaller than the symbol sizes.}
\label{fig:spin-log}
\vskip-2mm
\end{figure}

Gap information can also be extracted by a direct analysis of the large-$\tau$ 
decay of the correlation functions \cite{Sen2015,Suwa2016}. Considering the 
spin sector, the smallest singlet-triplet gap occurs at $\mathbf{q} = 
\mathbf{Q}$ and in the QD phase $S(\mathbf{Q},\tau)$ is dominated by the 
triplon mode. In the AFM phase, this gap corresponds to the lowest Goldstone 
mode, which has only a finite-size energy proportional to $1/N$. Thus 
$S(\mathbf{Q},\tau)$ decays very slowly with $\tau$ in this case and the 
dominant Goldstone contribution threatens to obscure the Higgs contribution 
\cite{Podolsky2011,Podolsky2012,Gazit2013,GazitPRB2013,Katan2015}. Examples 
of imaginary-time data for $S(\mathbf{Q},\tau)$ and of gap extractions are 
presented in Secs.~SII and SIII of the SM \cite{sm}.

We begin the discussion of our results by analyzing the triplon gap in the 
QD phase ($g > g_c$). For a given value of $g$, we extract the finite-size 
gap, $\Delta_T(L)$, from $S(\mathbf{Q},\tau)$ for a range of system sizes. As 
shown in Fig.~\ref{fig:spin-log}(a), $\Delta_T(L)$ decreases with increasing 
$L$ before converging to the thermodynamic limit. The extrapolated values of 
$\Delta_T(g)$ are shown in Fig.~\ref{fig:spin-log}(b) as a function of the  
separation ($|g - g_c|/g_c$) from the QCP. 

In the $\phi^4$ theory for an O($N$) order parameter, at $D = D_c$ one 
expects physical quantities to exhibit power-law scaling with mean-field 
critical exponents, but with multiplicative log corrections 
\cite{Zinn-Justin2002,Kenna2004292}, which have now been found in a 
number of recent studies \cite{Scammell2015,Qin2015,Jiang2016}. The 
scaling form of the triplon gap can be obtained directly from the 
correlation length ($\Delta \sim 1/\xi$), whence 
\begin{equation}
\Delta_T \thicksim (|g - g_c|/g_c)^{\nu} \ln^{-\hat\nu}(|g - g_c|/g_c),
\label{eq:lst}
\end{equation}
with $\nu = 1/2$ \cite{Zinn-Justin2002,sachdev2009exotic} and $\hat\nu = 
{(N+2)}/{2(N+8)}$ from perturbative renormalization-group calculations 
\cite{Kenna2004292,Kenna2012}, i.e.~$\hat\nu = {5}/{22} \approx 0.227$ for 
$N = 3$. It is clear from Fig.~\ref{fig:spin-log}(b) that Eq.~(\ref{eq:lst}) 
describes the data far better than the pure mean-field form and, by performing
an optimized fit \citep{Qin2015} with $\hat\nu$ as a free parameter, we deduce 
the exponent $\hat{\nu} = 0.230(2)$, fully consistent with the theoretical 
prediction.

\begin{figure}[t]
\vskip-1mm
\includegraphics[width=7.0cm]{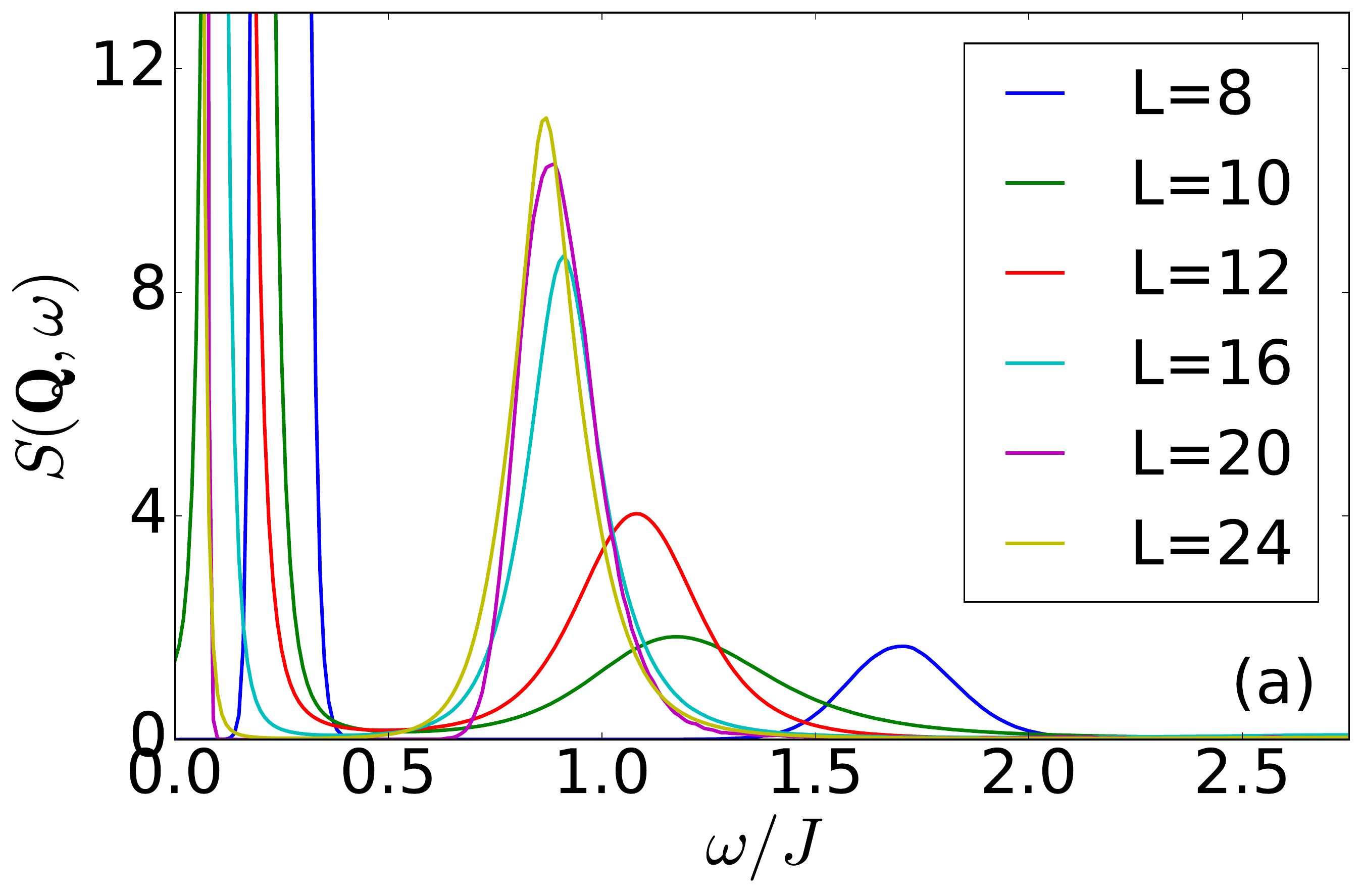}
\includegraphics[width=7.1cm]{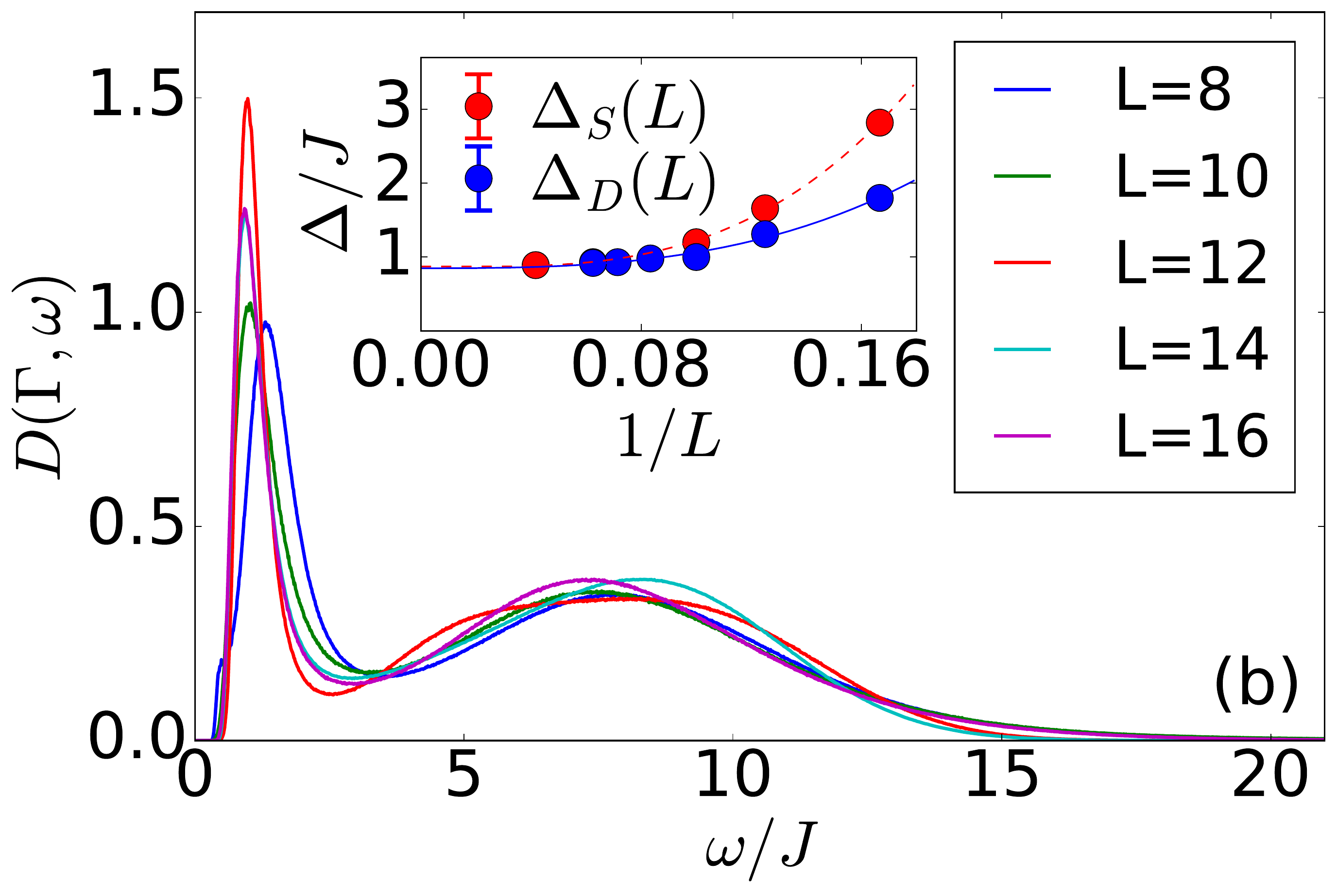}
\caption{(a) $S(\mathbf{Q},\omega)$ and (b) $D(\mathbf{\Gamma},\omega)$ 
obtained by SAC at $g = 4.724$ for different system sizes. The large 
low-energy (Goldstone) peak in panel (a) is cut off in order to show the 
secondary (Higgs) peak. The lower peak in panel (b) is the Higgs mode. The 
positions of both Higgs contributions converge with increasing $L$ to the 
same thermodynamic limit, as shown in the inset of panel (b). Spectral 
features outside the energy ranges shown are extremely weak.}
\label{fig:higgss}
\vskip-1mm
\end{figure}

To study the amplitude mode in detail, we analyze the spectral functions 
$S(\mathbf{Q},\omega)$ and $D(\mathbf{\Gamma},\omega)$ in the AFM phase 
($g < g_c$) near $g_c$. Figure \ref{fig:higgss} shows both quantities at 
$g = 4.724$ for several system sizes. Because SSE-QMC calculations of 
$D(\mathbf{\Gamma},\tau)$ are significantly more demanding (Sec.~SI), 
these are restricted to $L \le 16$, whereas for $S(\mathbf{Q},\tau)$ we 
access sizes up to $L = 24$. 

$S(\mathbf{Q},\omega)$ [Fig.~\ref{fig:higgss}(a)] is dominated by the 
Goldstone contribution, whose energy (spectral weight) is proportional to 
$1/N$ ($N$) at $T = 0$ (becoming the magnetic Bragg peak as $L \to \infty$). 
The Higgs spectral weight also diverges as $g \to g_c$; away from $g_c$ the 
Higgs mode remains as a clearly resolved finite-energy peak with convergent 
spectral weight, as also observed experimentally in TlCuCl$_3$ 
\cite{Ruegg2008,Merchant2014}. In $D(\mathbf{\Gamma},\omega)$ 
[Fig.~\ref{fig:higgss}(b)], the Higgs contribution is the distinctive 
low-energy peak. It is separated by a region of suppressed spectral weight 
from a broad maximum at higher energies due to multiple excitations. At low 
energies one expects a characteristic scaling form on which we comment in 
detail below. 

We observe good convergence with increasing $L$ in each of 
$S(\mathbf{Q},\omega)$ and $D(\mathbf{\Gamma},\omega)$. The peak widths in both 
quantities are invariant on increasing the amount of QMC data, demonstrating 
that any artificial broadening arising from the SAC procedure is negligible.
Examples of supporting tests are presented in Sec.~SII of the SM \cite{sm}. We 
have confirmed by a bootstrapping analysis that the fluctuations in the height 
and width of the lower $D(\mathbf{\Gamma},\tau)$ peak for $L \ge 10$ in 
Fig.~\ref{fig:higgss}(b) reflect statistical errors. Our system sizes are 
sufficient for a reliable study of the $L \to \infty$ limit in both sectors 
for the $g$ values shown in Fig.~\ref{fig:higgss} (i.e.~$g \approx g_c
 - 0.1$). 

We find that the positions of the finite-energy peaks in $S(\mathbf{Q},\tau)$ 
and $D(\mathbf{\Gamma},\tau)$ converge to the same value as $L \to \infty$ 
[inset, Fig.~\ref{fig:higgss}(b)]. In the phenomenological U(1) model for 
the broken-symmetry phase, one expects the $S = 0$ Higgs mode to be an 
elementary scalar \cite{Varma2015}, and thus in the AFM phase that the 
Higgs part of the $S = 1$ spectrum arises from a combination of this scalar 
with a gapless spin wave ($S = 1$, $\mathbf{q} = \pm \mathbf{Q}$). Although 
our finite-size calculations contain no explicit symmetry-breaking, they 
reflect this physics directly in that the spin peak lies higher than the 
dimer peak and their energy difference scales with $1/N$, as expected for 
a Goldstone mode. Thus the consistency between peaks in the $S = 0$ and 1 
spectral functions provides strong confirmation that both do indeed 
correspond to the Higgs mode. 

In Fig.~\ref{fig:spin-log}(b) we compare the extrapolated Higgs energies in 
the AFM phase with the triplet gaps in the QD phase at the same distance, 
$|g - g_c|/g_c$, from the QCP. The predicted $\sqrt{2}$ ratio 
\cite{sachdev2009exotic,Katan2015,Scammell2015} between $\Delta_{H}$ and 
$\Delta_{T}$ is clearly obeyed over this rather broad coupling range. We 
stress that this relation implies the presence of equivalent multiplicative 
log corrections [Eq.~(\ref{eq:lst})] to both $\Delta_T$ in the QD phase and 
$\Delta_H$ on the AFM side. 

\begin{figure}[t]
\vskip-1mm
\includegraphics[width=7.25cm]{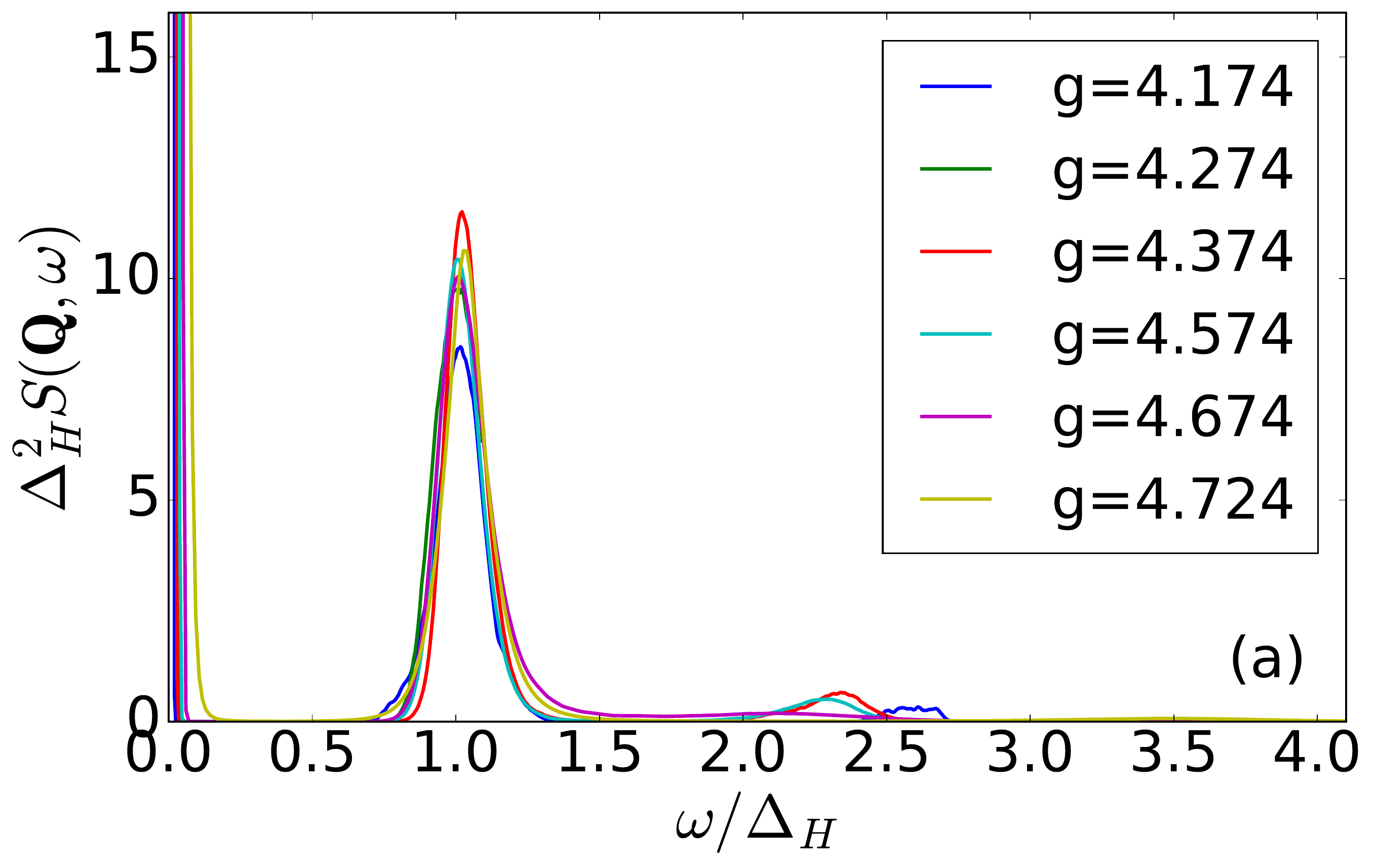}
\includegraphics[width=7.25cm]{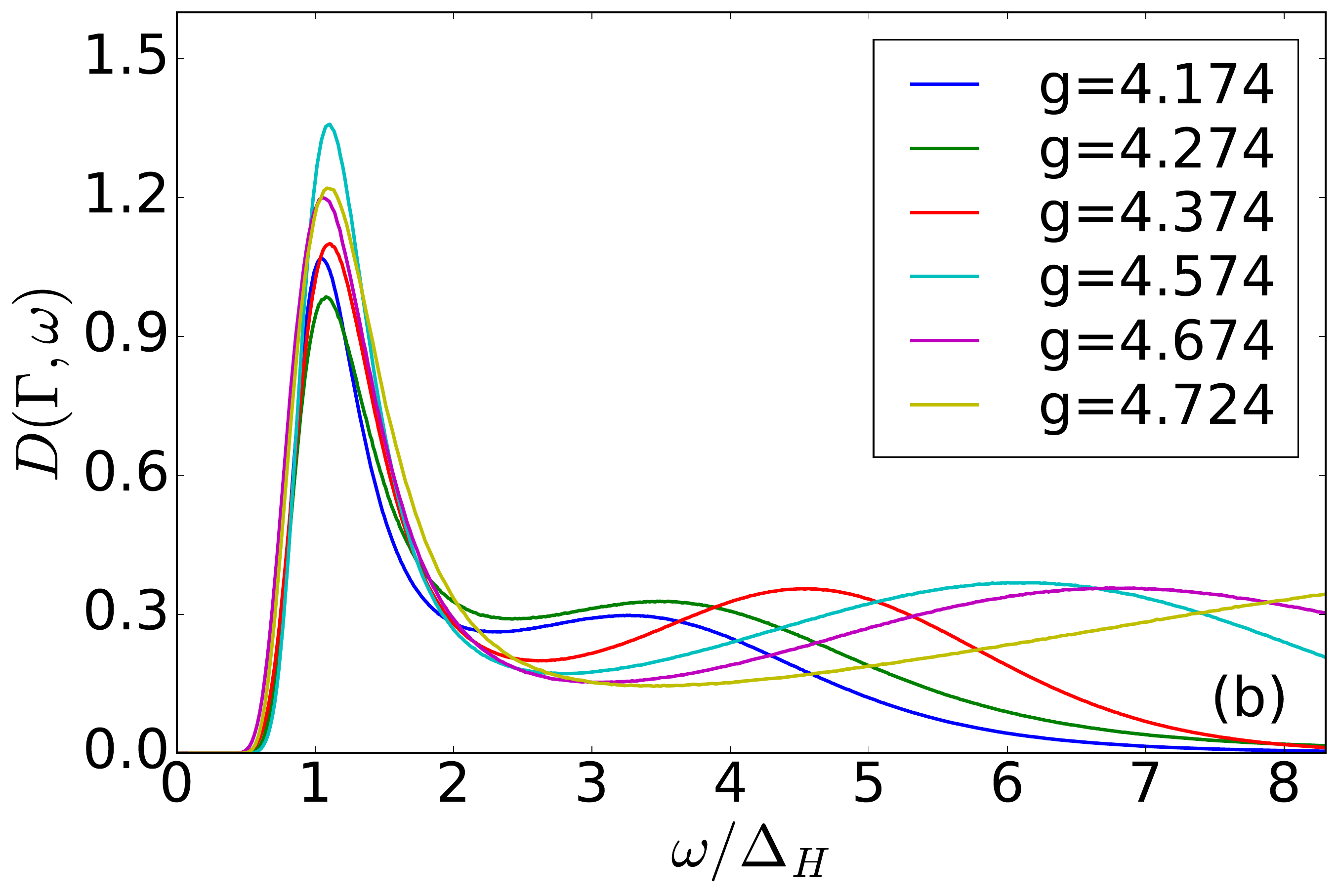}
\includegraphics[width=7.25cm]{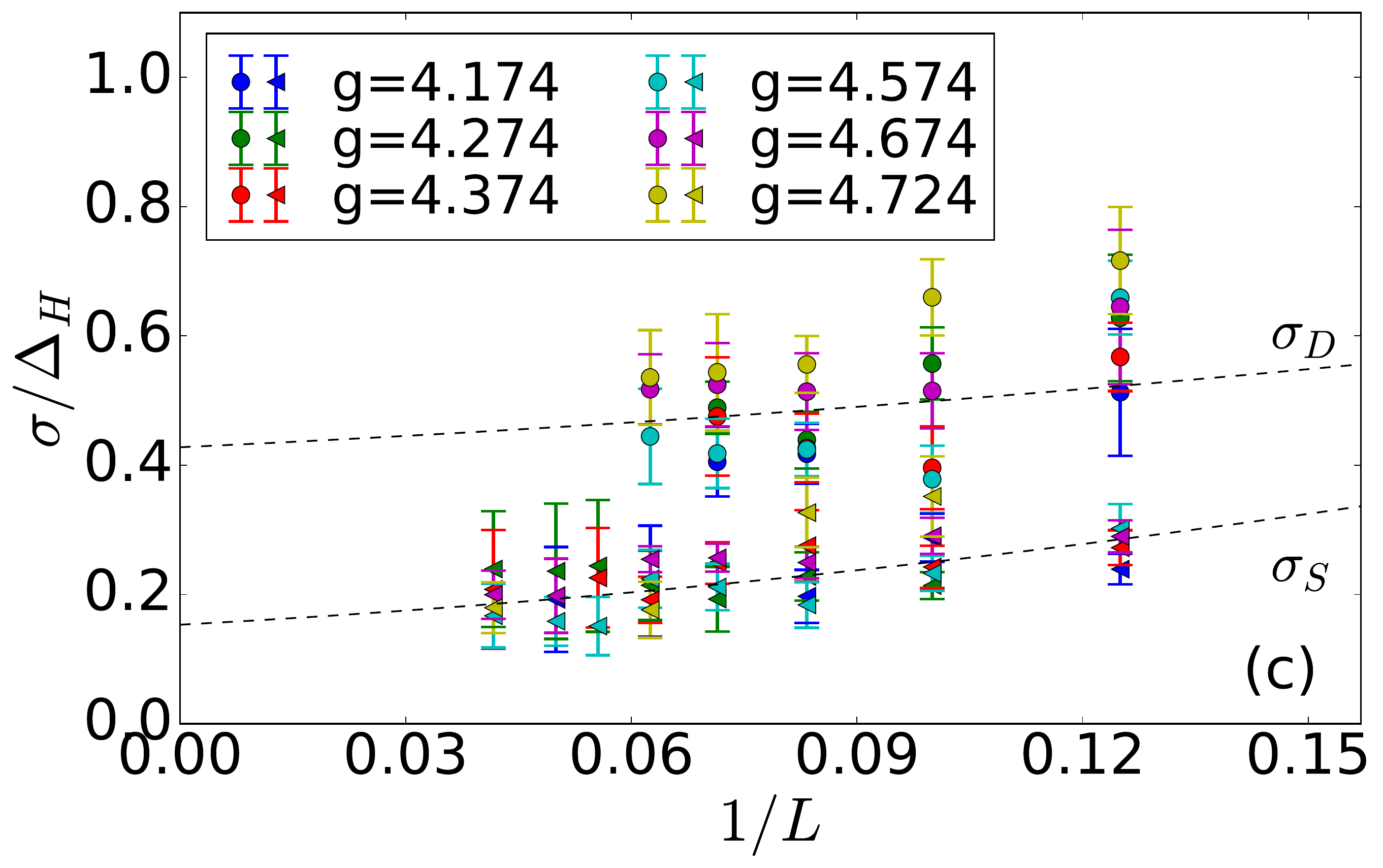}
\caption{(a) Scaled spectrum, $\Delta_H^2 S(\mathbf{Q},\omega/\Delta_H)$, 
calculated with $L = 24$ for a range of $g$ values. (b) $D(\mathbf{\Gamma},
\omega/\Delta_H)$ calculated with $L = 16$. (c) Width-to-energy ratios shown 
as functions of $1/L$. Circles and triangles are obtained respectively from 
$D(\mathbf{\Gamma},\omega/\Delta_H)$ and $S(\mathbf{Q},\omega/\Delta_H)$. 
Dashed lines are second-order polynomial fits to error-weighted average 
ratios.}
\label{fig:higgsw}
\vskip-2mm
\end{figure}

To investigate the scaling properties of the spectral functions near the QCP, 
we normalize $\omega$ by the $L \to \infty$ Higgs gap; results for $\Delta_H^2 
S(\mathbf{Q},\omega/\Delta_H)$ and $D(\mathbf{\Gamma},\omega/\Delta_H)$ are 
shown respectively in Figs.~\ref{fig:higgsw}(a) and \ref{fig:higgsw}(b) for 
the largest accessible system sizes. The amplitude-mode contributions to 
both the spin and dimer spectral functions exhibit near-ideal data collapse 
when scaled in this way. The collapse of the peak positions indicates that 
our data represent the quantum critical regime and the thermodynamic limit. 
The collapse of the peak widths demonstrates the critically damped nature 
of the Higgs mode. We note that Fig.~\ref{fig:higgsw}(a) also indicates 
the spectral weight of the next-order $S = 1$ processes, whose peak 
positions near $\omega = 2\Delta_H$ suggest excitations involving two Higgs 
modes, but statistical errors preclude a deeper analysis.

A universal scaling form for the scalar susceptibility (dimer spectral 
function) in the vicinity of the QCP,
\begin{equation}
D(\mathbf{\Gamma},\omega) \thicksim \Delta^{d+z-2/\nu}_{H} \Phi(\omega/\Delta_H),
\label{eq:collapse}
\end{equation} 
has been derived perturbatively in $1/N$ for the $O(N)$ model 
\cite{Podolsky2012,Gazit2013,GazitPRB2013} and by a $4 - \epsilon$ 
expansion \cite{Katan2015}. In (3+1)D with $z = 1$, one expects 
$D(\mathbf{\Gamma},\omega) = \Phi(\omega/\Delta_H)$, which is fully 
consistent with the data in Fig.~\ref{fig:higgsw}(b). This type of scaling 
has been documented in (2+1)D for both O(2) \cite{Pollet2012,Chen2013,
Gazit2013,GazitPRB2013,Rose2015} and O(3) models \cite{Rose2015,Wessel2015}, 
but Fig.~\ref{fig:higgsw}(b) constitutes the only unbiased numerical 
demonstration to date in (3+1)D. The infrared tail is expected 
\cite{Podolsky2011} to have the scaling form $D(\mathbf{\Gamma},\omega) 
\propto \omega^4$, but with the available system sizes is too weak to verify
this. For $S(\mathbf{Q},\omega/\Delta_H)$, we obtain data collapse by 
appealing to the result \cite{Ruegg2008} that the integrated spectral weight 
diverges as $1/\Delta_H$ when $g \to g_c$, which requires a rescaling by 
$\Delta_H^2$ [Fig.~\ref{fig:higgsw}(a)].

The scaling function $\Phi(\omega/\Delta_H)$ is shown in Ref.~\cite{Katan2015} 
to approach a $\delta$-function at $g = g_c$, due to the presence of log 
corrections in the width-to-energy ratios \cite{ra,raw}. For a quantitative 
analysis of the Higgs-peak widths, in Fig.~\ref{fig:higgsw}(c) we show the 
size-dependence of the ratios obtained from the FWHM $\sigma_S$ of the spin 
and $\sigma_D$ of the dimer peak. The error bars obtained by bootstrapping 
are significant, but it is clear that (i) the $L$-dependence of $\sigma_S/
\Delta_H$ and $\sigma_D/\Delta_H$ is weak, (ii) any $g$-dependence is weak, 
and (iii) $\sigma_D$ exceeds $\sigma_S$ by a factor of 3. We fit error-weighted 
averages of the width ratios, obtained from all $g$ values at each $L$, to a 
quadratic polynomial in $1/L$, as shown in Fig.~\ref{fig:higgsw}(c). At the 
mean-field level, we estimate the constant ratios ${\sigma_S}/{\Delta_H} = 
0.15(4)$ and ${\sigma_D}/{\Delta_H} = 0.43(6)$. The log dependence on 
$|g - g_c|$ is too weak to discern given the quality of the present data 
and the separation from the critical point. However, future calculations 
with smaller $|g - g_c|$, larger system sizes, and higher precision in 
$G(\tau)$ and $D(\tau)$ should be able to detect log corrections also in 
the width-to-energy ratios.

Remarkably, our SAC value for ${\sigma_S}/{\Delta_H}$ on the double-cubic 
lattice is in excellent agreement with the neutron scattering results for 
TlCuCl$_3$ near its QCP \cite{Ruegg2008,Merchant2014,Kulik11}. Given the 
difference in lattices and couplings, this result mandates a deeper 
investigation of possible reasons for a very weak dependence on microscopic 
details. The significantly larger value of ${\sigma_D}/{\Delta_H}$ reflects 
the different states probed by the two spectral functions, namely the 
elementary Higgs ($S = 0$) and combined Higgs--Goldstone ($S = 1$) 
excitations. This also implies different matrix-element effects in the 
peak shapes, which are evident in the different scaling forms of the peak 
areas in Fig.~\ref{fig:higgsw}. 

In summary, we have used large-scale quantum Monte Carlo simulations to 
investigate the quantum critical dynamics of the amplitude (Higgs) mode 
in a 3D dimerized antiferromagnet. Our work demonstrates that modern SAC 
methods are capable of resolving complex spectral functions, here with 
two peaks and non-trivial scaling behavior of both the peak widths and
heights. Our results not only verify the scaling predictions based on 
field-theory methods but also provide line-width information and 
nonuniversal factors that lie beyond current analytical treatments. 
The type of calculations reported here can be performed for different 
momenta, to study the dispersion of the amplitude mode and the evolution 
of its width in the spin and dimer sectors, as well as for the lattice 
geometry and exchange couplings of TlCuCl$_{3}$.

\begin{acknowledgments}
{\it Acknowledgments.}---We thank Stefan Wessel for communicating the results 
of a related investigation \citep{Wessel2016} prior to publication, Hui Shao 
for collaborations on the SAC method, and Oleg Sushkov for discussions. 
Numerical calculations were performed on the Tianhe-1A platform at the 
National Supercomputer Center in Tianjin. YQQ and ZYM acknowledge support from 
the Ministry of Science and Technology of China under Grant No.~2016YFA0300502, 
the National Science Foundation of China under Grant Nos.~11421092 and 
11574359, and the National Thousand-Young-Talents Program of China. YQQ 
would like to thank the Condensed Matter Theory Visitors Program of Boston 
University. AWS was supported by the NSF under Grant No.~DMR-1410126 and 
would also like to thank the Institute of Physics of the Chinese Academy of 
Sciences for visitor support.
\end{acknowledgments}



%

\newpage

\setcounter{page}{1}
\setcounter{equation}{0}
\setcounter{figure}{0}
\renewcommand{\theequation}{S\arabic{equation}}
\renewcommand{\thefigure}{S\arabic{figure}}

\begin{widetext}
\section*{Supplemental Material}

\centerline{\bf The amplitude mode in three-dimensional dimerized 
antiferromagnets} 
  
\vskip3mm

\centerline{Y. Q. Qin, B. Normand, A. W. Sandvik, and Z. Y. Meng}

\vskip6mm

Here we provide the technical details concerning our computations of 
imaginary-time spin and dimer correlation functions (Sec.~SI), of the 
stochastic analytic continuation method for extracting the full real-frequency 
spectral functions (Sec.~SII), and of gap extraction from the decay of the 
measured correlation functions at long imaginary times (Sec.~SIII).

\end{widetext}

\subsection{SI. Evaluation of dynamical correlation \\ functions in SSE-QMC}

We wish to compute the product of two operators, one of which is displaced 
in imaginary time, 
\begin{equation}
\hat{O}_{2}(\tau) \hat{O}_{1}(0) = e^{\tau\hat{H}} \hat{O}_2 e^{-\tau\hat{H}} \hat{O}_1.
\end{equation} 
Within SSE-QMC~\cite{Sandvik1991,Sandvik1992,Sandvik1999,Syljuaasen2002}, the 
exponential functions are Taylor-expanded, whereupon the time displacement
is converted into a sum over the separations $m$ between states propagated 
by $m$ operators in the importance-sampled operator strings originating from 
the Taylor expansion of the density matrix, ${\rm exp}({-\beta H})$. Here we 
present the relevant formulas in order to make some comments concerning the
implementation of the method, and refer to the papers cited above for details 
of the SSE scheme and the origin of these formulas.

If the operators $\hat{O}_{1}$ and $\hat{O}_{2}$ are terms of the Hamiltonian
(for example the dimer operators in the present study, which are Heisenberg
exchange operators), for a given operator string one has \cite{Sandvik1992}
\begin{widetext}
\begin{equation}
\langle \hat{O}_1(\tau) \hat{O}_2(0) \rangle = \left\langle \sum_{m=0}^{n-2} 
\frac{\tau^{m} (\beta - \tau)^{n-m-2}}{\beta} \frac{(n-1)!}{(n-m-2)!m!} 
N(O_1,O_2;m) \right\rangle,
\label{eq:dcg}
\end{equation}
where $n$ is the expansion order of the SSE configuration and $N(O_1,O_2;m)$ 
is the number of times the two operators are separated by $m$ other
operators in the string. If the two operators are diagonal in the basis of 
the SSE expansion, the correlation function can be written as
\begin{equation}
\langle \hat{O}_1(\tau) \hat{O}_2(0) \rangle = \left\langle \sum_{p=0}^{n-1} 
\sum_{m=0}^{n} \frac{\tau^{m} (\beta - \tau)^{n-m}}{\beta^{n}} \frac{(n-1)!}
{(n-m)!m!} O_1(\vert p \rangle) O_2(\vert p+m \rangle) \right\rangle,
\label{eq:dcd}
\end{equation} 
\end{widetext}
where $\vert p \rangle$ is the state propagated by $p$ operators relative 
to a stored SSE state.

In the above expressions for the correlation functions, the prefactors of 
$N(O_1,O_2;m)$ or $O_1(\vert p \rangle) O_2(\vert p+m \rangle)$ are strongly 
peaked at $m \approx n\tau/\beta$,
and it is therefore not necessary to evaluate the full sums over $m$; for a 
given $\tau$, a pre-computed $m$-range of sufficient width, meaning one 
containing approximately $\sqrt{n}$ operators, can be used. Nevertheless, 
these summations are time-consuming and can often dominate the computational 
effort. In Ref.~\cite{Wessel2015}, a modified kernel was introduced in the 
analytic continuation to allow a different discretization of the correlation 
function (\ref{eq:dcg}) used for the dimer correlations. Here we choose to 
use the full form, thus avoiding a potential loss of frequency resolution 
due to the further convolution implied within the modified kernel. In order 
to ensure that the most time-consuming measurements are not carried out more 
often than necessary, by minimizing autocorrelations, we separate our 
measurements of the dimer-dimer correlation function by $L$ SSE update sweeps.

Because spin-spin correlations are diagonal in the SSE basis, we can adopt a 
different and faster approach than the form (\ref{eq:dcd}). We use a 
time-sliced form of the density matrix,
\begin{equation}
{\rm e}^{-\beta H} = \prod\limits_{l=1}^M {\rm e}^{-\Delta H},~~~\Delta = \beta/M,
\end{equation}   
and carry out the SSE simulations with each of the time slices, 
${\rm e}^{-\Delta H}$, expanded individually. This allows easy access to all 
correlation functions of diagonal operators separated by times that are 
multiples of the step $\Delta$, simply by measuring the correlations in the 
propagated states at the boundaries between time slices.

In most cases we use a quadratic time grid, instead of the uniform grid 
normally used, with the times in the set $\{\tau_i\}$ of the form
$\tau_i = \Delta i^2$ for $i = 0, 1, 2, \ldots$ This type of grid is useful 
when the inverse temperature, $\beta$, is large, as it gives access to both 
short and long time scales with a reasonably small set of time points 
(typically numbering in tens rather than hundreds) for which the covariance 
matrix \cite{JARRELL1996133} required in analytic continuation can be computed 
properly. In Fig.~\ref{fig:spin-correlation} we show examples of the spin 
correlation functions in the QD and AFM phases, whose analysis we discuss 
in Secs.~SII and SIII.

\section{SII. Stochastic analytic continuation}
\label{sm:sac}

The imaginary-time correlation function of an operator $\hat O$, 
$G(\tau) = \langle \hat O(\tau) \hat O(0)\rangle$, is related to a
corresponding spectral function, $A(\omega)$, according to
\begin{equation}
G(\tau) = \int_{-\infty}^\infty d\omega A(\omega) K(\tau,\omega),
\label{gtaurel1}
\end{equation}
where the kernel, $K(\tau,\omega)$, depends on the type of the spectral 
function. We wish to deduce the spectral function normally referred to as 
the dynamic structure factor (for the spin or dimer operator, as defined in 
the main text), which has the spectral representation 
\begin{equation}
A(\omega) = \frac{1}{\pi} \sum_{n} {\rm e}^{-\beta E_n} \sum_{m} |\langle m| 
\hat O|n\rangle|^2 \delta(\omega - E_n + E_m)
\end{equation}
in terms of the energy eigenvalues and the corresponding matrix elements of
the operator $\hat O$. 

For the bosonic case studied here, the spectra at positive and negative 
frequencies obey the relation $A(-\omega) = {\rm e}^{-\beta\omega} A(\omega)$,
whence one may restrict the integral in Eq.~(\ref{gtaurel1}) to positive 
frequencies by using the kernel 
\begin{equation}
K(\tau,\omega) = \frac{1}{\pi} (e^{-\tau\omega} + e^{-(\beta-\tau)\omega}).
\label{eq:kernel1}
\end{equation}
The normalization of the spectrum can then be expressed as 
\begin{equation}
G(0) = \int_{0}^\infty d\omega A(\omega) K(\tau=0,\omega).
\end{equation}
In the version of the SAC method used here \cite{Shao2016}, we adopt the 
normalization $G(0) = 1$ [and later multiply the final spectrum by the 
original value of $G(0)$] in order to work with a spectral function that is 
itself normalized to unity on the positive frequency axis, i.e.~without being 
multiplied by the kernel (\ref{eq:kernel1}). We therefore define
\begin{equation}
B(\omega) = A(\omega)(1 + {\rm e}^{-\beta\omega}),
\label{barelation}
\end{equation}
with normalization
\begin{equation}
\int_{0}^\infty d\omega B(\omega) = 1.
\end{equation}
Using a modified kernel,
\begin{equation}
\bar K(\tau,\omega) = \frac{1}{\pi}\frac{e^{-\tau\omega} + e^{-(\beta-\tau)\omega}}{1
 + e^{-\beta\omega}},
\label{eq:kernel2}
\end{equation}
the relationship between $B(\omega)$ and $G(\tau)$, normalized such that 
$G(0) = 1$, is
\begin{equation}
G(\tau) = \int_{0}^\infty d\omega B(\omega) \bar K(\tau,\omega).
\label{eir}
\end{equation}
Working with a set $\lbrace \tau_i \rbrace$ of imaginary-time points, for which 
the values $G(\tau_i)$ and the accompanying covariance matrix, $C_{ij}$, have 
been computed using QMC simulations, our aim is to invert the integral relation 
(\ref{eir}) numerically to acquire $B(\omega)$, and then to obtain $A(\omega)$ 
from Eq.~(\ref{barelation}).

\begin{figure}[t]
\includegraphics[width=7.5cm]{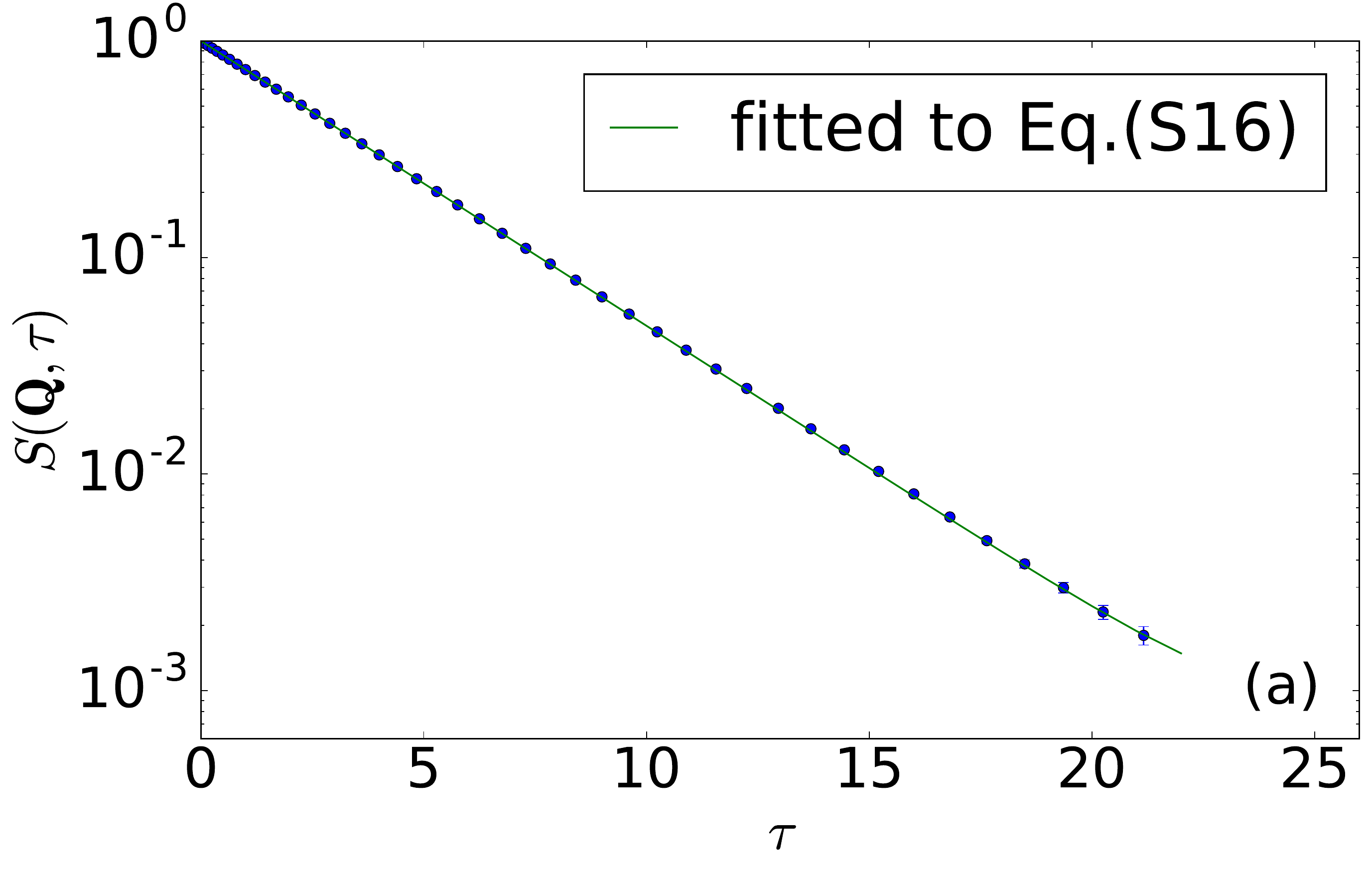}
\includegraphics[width=7.5cm]{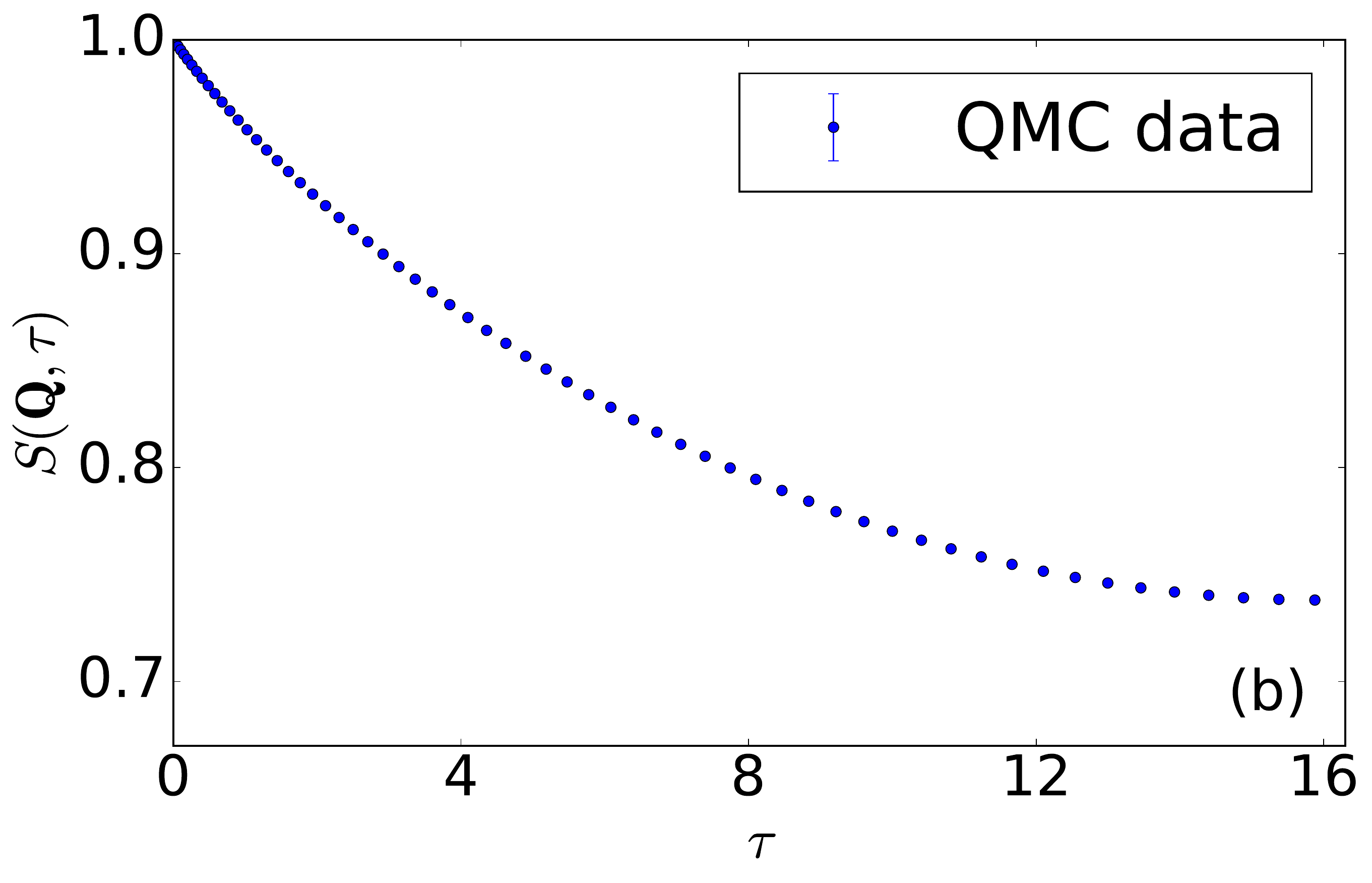}
\caption{Dynamical spin correlation functions for double-cubic systems with
(a) $L = 24$ in the QD phase at $g = 4.87$ and (b) $L = 16$ in the AFM 
phase at $g = 4.724$. A quadratic grid of imaginary times, $\tau_i = \Delta 
i^2$, $i = 0, 1, 2, \ldots$, was used in both cases and the inverse temperature 
was $\beta = 2L$. In panel (a), an almost pure exponential decay is observed 
and the gap is readily extracted from Eq.~(\ref{eq:fit_peak}) of Sec.~SIII, 
giving $\Delta_T = 0.3028(1)$ (with the error bar estimated by bootstrapping). 
The spectral functions obtained by analytic continuation are shown in the 
corresponding panels of Fig.~\ref{fig:spin-spectrums}.}
\label{fig:spin-correlation}
\end{figure}

To carry out the analytic continuation, we use a variant \cite{Shao2016} of 
the SAC approach 
\cite{Sandvik1998,Beach2004,Syljuasen2008,Fuchs2010,Sandvik2015}, where 
a chosen parameterization of the spectrum, typically using a large number of 
$\delta$-functions, is sampled in a Monte Carlo simulation using
a likelihood function 
\begin{equation}
P(B) \propto \exp(-\chi^2[B]/2\Theta),
\end{equation}
where $\Theta$ is a fictitious temperature and thus $\chi^2$ plays the role 
of an energy if the system is regarded as a problem in statistical mechanics.
It has been demonstrated \cite{Beach2004} that when this formulation is 
treated in a mean-field-type approximation, it reduces to the more commonly 
used Maximum Entropy (ME) method \cite{JARRELL1996133}. In general, one 
expects SAC to resolve finer details of the spectrum than ME, because it 
takes into account fluctuations around the mean-field solution, at the price 
of requiring long sampling times in certain cases (Refs.~\cite{Fuchs2010} 
and \cite{Bergeron2016} contain some discussion of this point). The most 
recent variant of the method \cite{Shao2016}, which we deploy here, reduces 
these sampling times considerably. Typically it delivers simple, single-peak 
spectral functions in a few minutes or less, when running on a single core, 
and more complicated spectra, such as the cases with two well-separated peaks 
that we consider here, in under an hour.

The sampling space is a large number, $N_\omega$, of movable $\delta$-functions 
placed on a frequency grid with a spacing, $\Delta_\omega$, sufficiently fine as 
to be regarded in practice as a continuum (e.g.~$\Delta_\omega = 10^{-5}$). 
$N_\omega$ may range from a value of order $100$ for spectra with little 
structure to $10^4$ or more for complicated spectra. The optimal value also 
depends on the quality (statistical errors) of the QMC data, with better data 
requiring larger $N_\omega$ for efficient sampling.

At a given instant, the frequency of the $i$th $\delta$-function is $\omega_i$, 
which is an integer multiple of $\Delta_\omega$. A sweep consisting of $N_\omega$ 
moves of the type $\omega_r \to \omega'_r$ is performed, with $r$ chosen 
randomly and the size of the change generated randomly within a window 
such that approximately half of the updates are accepted. The spectrum is 
accumulated in a histogram whose bin size is typically much larger than 
$\Delta_\omega$, and is chosen such that all features of the spectrum can be 
clearly resolved.

In the simplest case, the amplitudes of the $\delta$-functions are all 
the same, $b_i = 1/N_\omega$, with their value corresponding to a spectrum 
$B(\omega)$ normalized to unity, as discussed above. We stress that these 
amplitudes do not have to be changed, and a peak in the spectrum corresponds 
only to a high average density of $\delta$-functions within a corresponding 
region. We also carry out simultaneous updates of more than one 
$\delta$-function, a technique we will discuss in detail elsewhere 
\cite{Shao2016}. The constant amplitudes are intended to reduce the amount 
of entropic bias \cite{Sandvik2015} in the spectrum, and tests indicate that 
this development does indeed improve the fidelity of the method.

A further refinement is not to use the same amplitude for all 
$\delta$-functions, but to generate a range of varying amplitudes, 
such as $b_i \propto i$ (with a prefactor chosen to satisfy the standard 
normalization), and this is the method we employ throughout the current 
analysis. These amplitudes are kept constant during the sampling process and, 
because there are no constraints on the locations of the $\delta$-functions,
small amplitudes can migrate to areas with lower spectral weight. As long as 
$N_\omega$ is sufficiently large,  we find no differences in the final 
averaged spectrum between the methods of uniform or varying amplitudes. 
However, the sampling efficiency is increased in the method of varying
amplitudes, especially if we include simultaneous amplitude updates of two 
$\delta$-functions, of the type $b'_i = b_j$, $b'_j = b_i$, with $i$ and $j$ 
chosen randomly among all the $\delta$-functions. When sampling spectra with 
two or more peaks, this update leads to considerable improvements in the 
efficiency of transferring weight between peaks.

An important issue in SAC is how to select the sampling temperature, $\Theta$. 
The general situation is that a very low $\Theta$ freezes the spectrum close
to a stable or metastable $\chi^2$ minimum, while a high $\Theta$ leads to 
large $\chi^2$ values, i.e.~poor fits to the QMC data for $G(\tau)$. There is 
a range of $\Theta$ over which the average $\chi^2$ value is small but the 
fluctuations are significant, and cause a smoothing of the spectrum. There 
is no agreement on exactly how $\Theta$ should be chosen, but in general the 
different schemes proposed in the literature produce very similar results. 
This issue is similar to the various ways in which the entropic weighting of 
the spectrum can be chosen in the ME method \cite{JARRELL1996133,Bergeron2016}.

Here we adopt a simple temperature-adjustment scheme devised in 
Ref.~\cite{Shao2016}, where a simulated annealing procedure is first 
carried out to find the minimum value, $\chi^2_{\rm min}$ (in practice this is 
actually a value close to the minimum, as it may be very difficult to reach 
the exact global minimum of $\chi^2$). After this initial step, $\Theta$ is 
adjusted so that the average $\chi^2$ during the final sampling process for 
collecting the spectrum satisfies the criterion
\begin{equation}
\langle \chi^2 \rangle \approx \chi^2_{\min} + \sqrt{2N_\tau},
\label{eq:chi^2}
\end{equation}
where $N_\tau$ is the number of time points in the QMC dataset for $G(\tau)$. 
With $N_\tau$ replaced by $N_{\rm dof}$ (the number of degrees of freedom in a 
simple fitting procedure), the square-root term would be exactly the standard 
deviation of the $\chi^2$ distribution. With the spectrum constrained to 
positive frequencies, the number of degrees of freedom is not simply the 
difference $N_\tau - N_\omega$, because the sampling parameters, which here 
are the $N_\omega$ frequencies, cannot be considered as independent variables 
(we note that normally $N_\omega \gg N_\tau$, whence $N_{\rm dof}$ from the naive 
definition would be negative and therefore meaningless). We therefore use 
instead the quantity $\sqrt{2N_\tau}$, which typically will be one to two 
standard deviations of the distribution. This level of noise then corresponds 
roughly to the removal of distortions due to ``fitting to the statistical 
errors.'' In practice, we find that the criterion (\ref{eq:chi^2}) produces 
excellent spectra in tests both on synthetic imaginary-time data and on 
actual QMC results for systems with known spectral functions \cite{Shao2016}.

\begin{figure}[t]
\includegraphics[width=7.5cm]{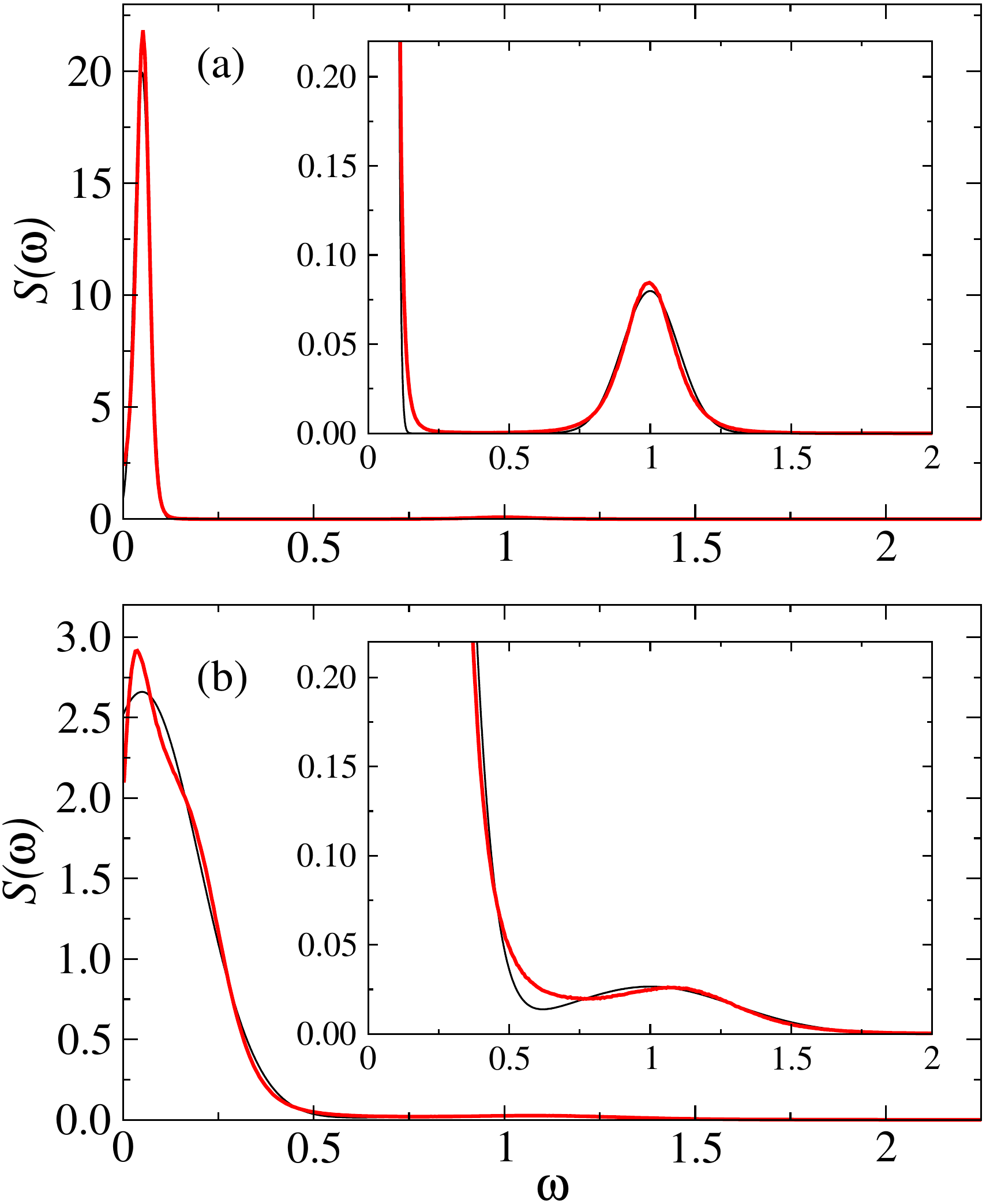}
\caption{Tests of the advanced SAC method performed using synthetic QMC data. 
Black curves show artificial spectral functions consisting of two Gaussian 
peaks. Red curves show the results obtained by applying SAC to the 
corresponding Green function, $G(\tau)$, with statistical noise generated 
as described in the text.}
\label{fig:sac-test}
\end{figure}

To illustrate this point, here we show two examples of the results obtained 
by applying SAC to synthetic QMC data. Our aim is to demonstrate the ability 
of the advanced SAC method to reproduce a very small secondary peak in the 
spectrum of precisely the type presented by the amplitude-mode contribution 
in Figs.~3(a) and 4(a) of the main text. We construct spectra consisting 
of two Gaussians, with a dominant narrow one at low frequency, containing 
98\% of the spectral weight, and a weak secondary one at higher frequencies. 
We consider two different cases. In Fig.~\ref{fig:sac-test}(a), the secondary 
peak is also relatively narrow and is separated from the dominant peak by a 
real gap (a region of zero spectral weight), while in 
Fig.~\ref{fig:sac-test}(b) the second peak is broad and there is no gap 
between the two peaks, but only a region of suppressed spectral weight.

To mimic the QMC data underlying the spectra shown in Figs.~3(a) and 4(a) of 
the main text, we set the inverse temperature to $\beta = 50$, use a similar 
number of $\tau$ points ($N_\tau \approx 30$), and generate a noise term, 
$\sigma_i$, to add to the imaginary-time correlation function, $G(\tau_i)$, 
corresponding to the artificial spectrum. Given that the QMC data for 
different $\tau$ points are strongly correlated, we construct correlated 
noise by averaging independently generated, normally-distributed noise data, 
$\sigma^0_i$, using an exponentially decaying weight function,
\begin{equation}
\sigma_i = \sum_{j} \sigma^0_i {\rm e}^{-|i-j|/\xi_\tau},
\end{equation}
which we add to $G(\tau_i)$. We generate a large number of such noisy data 
sets and run these through the same bootstrapping code that we use for the 
real QMC data shown in the main text. The correlation time, $\xi_\tau$, and 
the variance of $\sigma^0_i$ (which is the same for all $i$), are adjusted 
so that the eigenvalues of the covariance matrix are similar to those of 
our typical QMC data. 

Figure \ref{fig:sac-test}(a) shows that the present SAC procedure is fully 
capable of resolving the secondary peak in the spectrum when there is a gap, 
with minimal broadening and the correct peak shape. When there is no gap 
[Fig.~\ref{fig:sac-test}(b)], the overall shape of the spectrum, including 
the second peak, is also well reproduced; although the finite spectral weight 
between the two peaks in this case is larger than in the original spectrum 
[inset, Fig.~\ref{fig:sac-test}(b)], decreasing the noise level in the data 
improves this fit. It is clear from Fig.~\ref{fig:sac-test} that advanced 
SAC methods of the type we apply here are well able to resolve the types 
of spectral functions expected on physical grounds in all parameter 
regimes of the system discussed in the main text.

\begin{figure}[t]
\includegraphics[width=7.5cm]{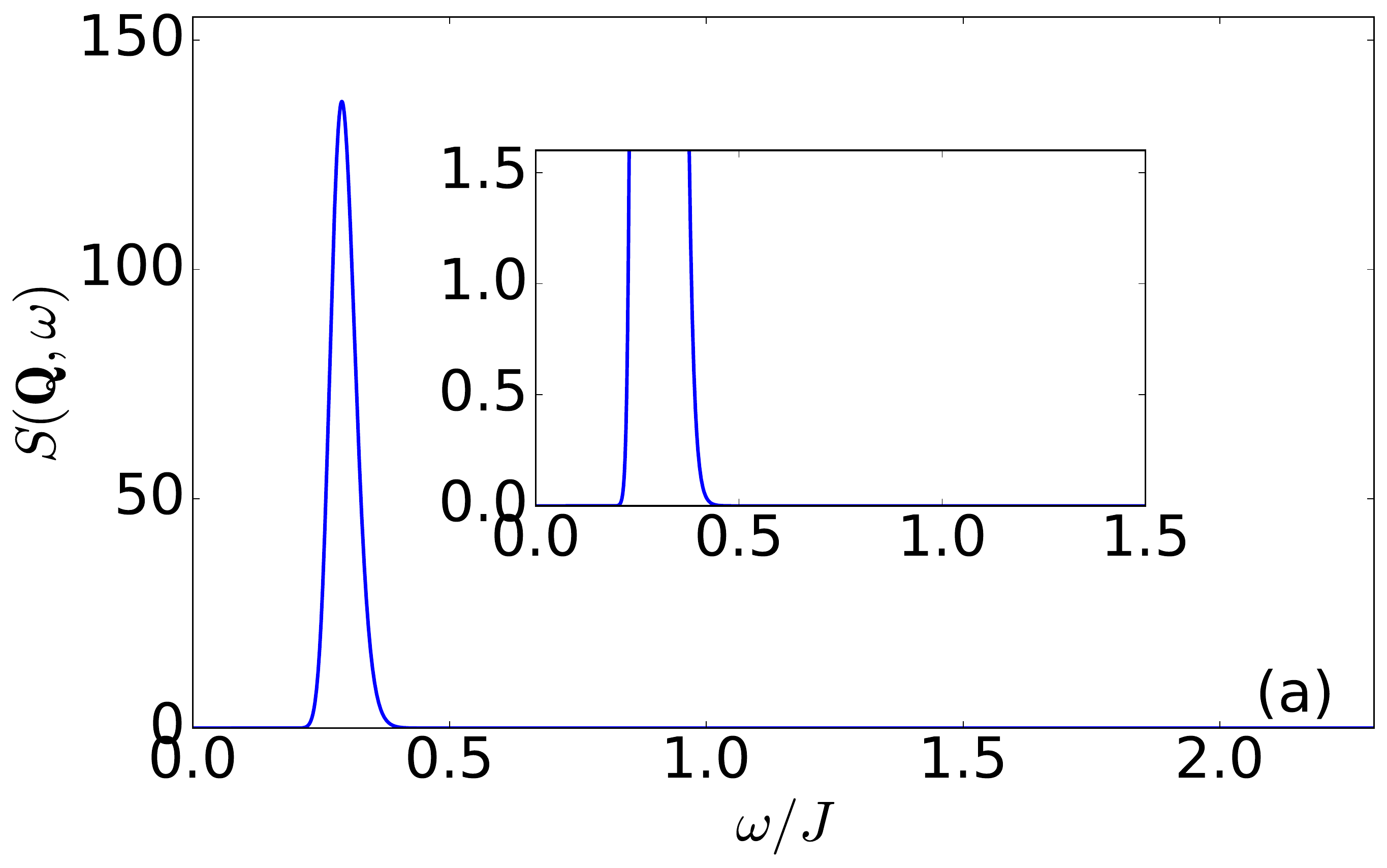}
\includegraphics[width=7.5cm]{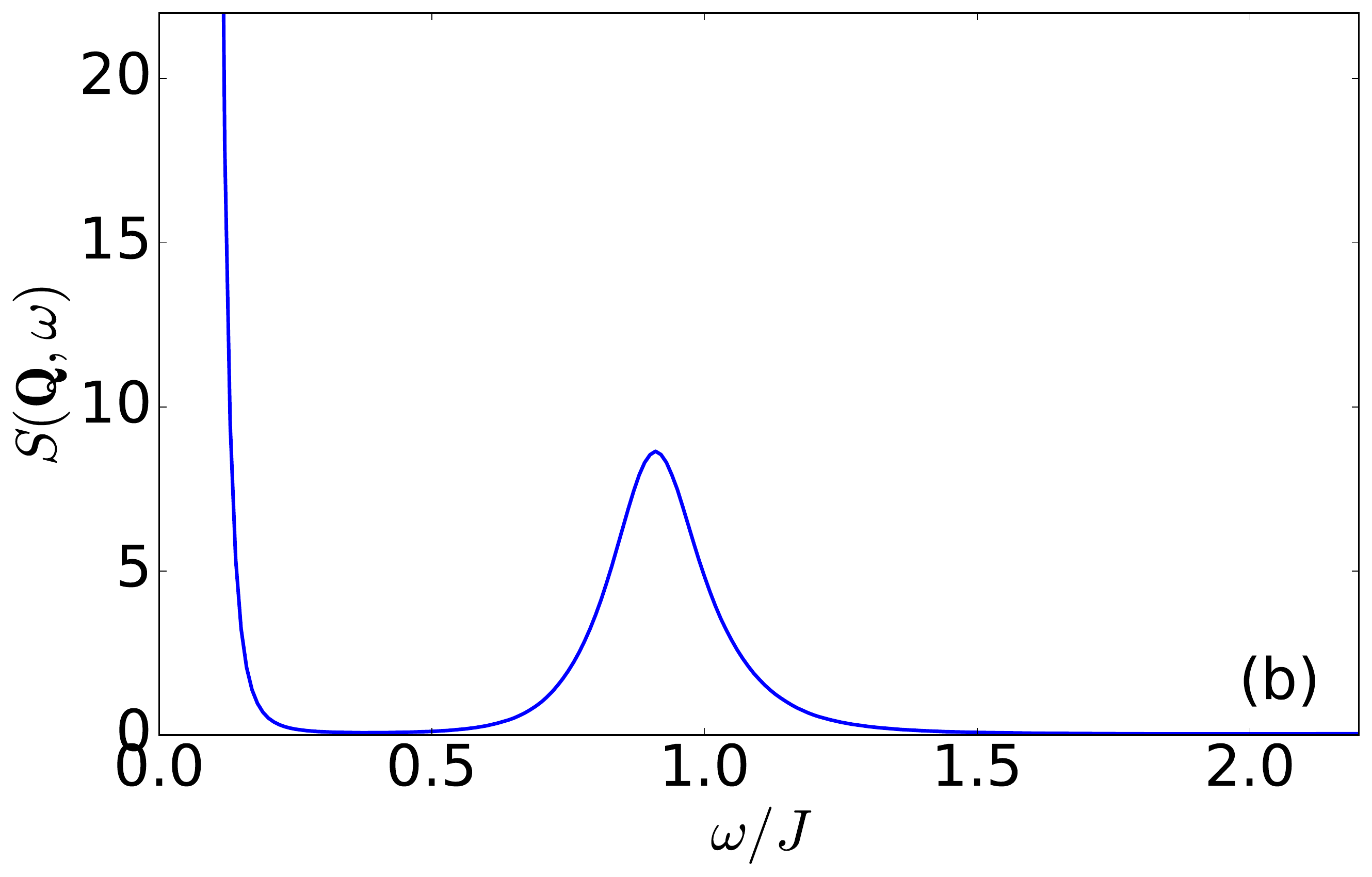}
\caption{Spectral functions obtained using the SAC method with 
the imaginary-time data shown in the corresponding panels of 
Fig.~\ref{fig:spin-correlation}. There is no significant spectral
weight beyond the frequency ranges shown.}
\label{fig:spin-spectrums}
\end{figure}

With this demonstration in hand, we conclude this section by showing the 
spectral functions obtained by applying our SAC method to the imaginary-time 
data in Fig.~\ref{fig:spin-correlation}. In Fig.~\ref{fig:spin-spectrums}(a), 
where the system is in the QD phase, only a single narrow triplon peak is 
found, corresponding to the essentially pure exponential decay seen in
Fig.~\ref{fig:spin-correlation}(a). By contrast, when the system is in the 
AFM phase, we observe the two peaks shown in Fig.~\ref{fig:spin-spectrums}(b),  
namely the Goldstone mode at $\omega \propto 1/N$ and a smaller peak at higher 
frequency that corresponds to the amplitude mode, whose scaling properties are 
discussed in the main text.

\section{SIII. Dominant-mode fitting}

If a spectral function, $A(\omega)$, contains a single $\delta$-function, 
the corresponding imaginary-time correlation function, $G(\tau)$, decays
exponentially at $T = 0$, with the decay constant being the inverse of 
the gap, $\Delta$. At a finite inverse temperature, $\beta$, the symmetry
$G(\beta - \tau) = G(\tau)$ of a bosonic function implies the form
\begin{equation}
\label{eq:fit_peak}
G(\tau) = a \cosh[ (\beta/2 - \tau) \Delta],
\end{equation}
where $a$ is a constant.

Even beyond the single-mode case, this form can in principle always be 
used to extract the gap of a finite system, whose spectrum is a sum of 
$\delta$-functions, by fitting to the long-time part of $G(\tau)$, where
the lowest-energy $\delta$-function dominates the decay. If the spectral 
weight of the lowest $\delta$-function, which by definition is exactly at 
the gap, $\Delta$, is sufficiently large and the second gap to the following 
$\delta$-function is also sufficiently large, then the asymptotic form of 
$G(\tau)$ is also given by Eq.~(\ref{eq:fit_peak}) if $\beta$ is sufficiently 
large. The gap may then be extracted by fitting the large-$\tau$ data. In 
practice, these conditions can be hard to fulfil rigorously and it becomes 
necessary to go beyond the single-mode form by performing a fit to more than 
one exponential, or to perform a still more sophisticated analysis of the 
long-time decay of the correlation functions. Some methods for this task 
have been discussed recently \cite{Sen2015,Suwa2016}.

Figure \ref{fig:spin-correlation} shows two examples of spin 
correlation functions from the double-cubic Heisenberg model. In 
Fig.~\ref{fig:spin-correlation}(a), the system is well inside the QD 
phase and the spectrum at ${\bf q} = {\bf Q} = (\pi,\pi,\pi)$ is completely 
dominated by a single triplon mode. In this case it is easy to extract 
the gap by fitting to Eq.~(\ref{eq:fit_peak}) and the results are 
identical to those obtained by SAC [Fig.~\ref{fig:spin-spectrums}(a)]. 
By contrast, in Fig.~\ref{fig:spin-correlation}(b) the system lies in 
the AFM phase, close to the QCP, where the resulting spectral function 
contains both the Goldstone mode, appearing at an energy proportional 
to $1/N$ and with significant finite-size broadening, and the weak 
amplitude-mode contribution at higher energies. In this case it is very 
difficult to extract any information by single-mode fitting and a different 
technique is required to obtain $A(\omega)$; Fig.~\ref{fig:spin-spectrums}(b) 
shows the results obtained by using SAC for this purpose.

\end{document}